\documentclass[12pt]{article}
\usepackage{amssymb,amsfonts}
\usepackage{epsf,epsfig}
\begin{document}
\baselineskip 18pt
\title
{Zero-temperature magnetic-field-induced phase transition between two ordered
gapped phases in spin-ladders with ferromagnetic legs}
\author{P.~N.~Bibikov}
\date{\it V.~A.~Fock Institute of Physics,\\ Sankt-Petersburg State
University, Russia}

\maketitle

\vskip5mm

\begin{abstract}
We suggest that under an increase of magnetic field a spin-ladder
with ferromagnetic legs does not pass without fail through an
incommensurate phase but possibly in a straight way turns into a
fully polarized ferromagnetic phase. The spin gap remains finite
at the transition point. This scenario of zero-temperature first order
phase transition is demonstrated for two special solvable spin-ladder models.
\end{abstract}

\section{Introduction}
Low-temperature magnetic phase transitions in spin-ladders were
intensively studied in the last decade both theoretically
\cite{1}-\cite{8} and experimentally \cite{8}-\cite{10}. In all
these papers (even in the Ref. \cite{6} devoted especially to
magnetic behavior of spin-ladders with ferromagnetic legs) the
theoretical scenario was quite identical. At zero magnetic field
the ladder has a non-magnetic gapped ground state. An increase of
magnetic field entails a decrease of the gap. At the critical
value of the magnetic field $h_c$ the gap closes and the system
turns into a gapless incommensurate magnetic phase. A further
increase of the magnetic field from $h_c$ up to the saturation
field $h_s$ entails a continuous change of the gapless ground
state and an increase of magnetization up to the maximum saturated
value corresponding to the full polarization. Any further increase
of the magnetic field does not change the vacuum entailing only an
appearance and increase of a gap.

This picture is based on the following theoretical argumentation \cite{1}-\cite{8}.
In zero magnetic field the ladder has the factorized singlet-rung
ground state:
\begin{equation}
|vac\rangle_0=\prod_n|0\rangle_n,
\end{equation}
where $|0\rangle_n$ is the singlet state associated with $n$-th
rung. All low-lying excitations originate from sparse
rung-triplets in the rung-singlet sea. Induced by magnetic field
Zeeman splitting entails the decrease of the gap with the rate
proportional to the total spin of the state. The gaps
corresponding to high-spin sectors decrease more rapidly, however
the low-spin sectors have more narrow gaps. By the latter reason
it was always assumed in \cite{1}-\cite{10} that the one-magnon gap
closes faster than the multi-magnon ones. Under this assumption
the critical value of magnetic field may be expressed from the
one-magnon spin gap according to the formula,
\begin{equation}
g\mu_Bh_c=E_{gap}^{magnon}.
\end{equation}

However as it will be shown below these arguments fail for the
spin-ladders with ferromagnetic legs. Really let us first consider
a spin-ladder with zero coupling along the legs or equivalently a
set of isolated dimers. For this system $h_c=h_s$. This means that
at $h=h_c$ the system turns into the fully polarized ferromagnetic
ground state,
\begin{equation}
|vac\rangle_s=\prod_n|1\rangle_n^1.
\end{equation}
Here $|1\rangle^j_n$ ($j=-1,0,1$) is the $S=1$ triplet associated
with $n$-th rung. An addition of a ferromagnetic coupling along
legs decreases the energy of the fully polarized state and prompts
it to reach the zero-energy level first. The system will turn
into the state (3) escaping a commensurate-incommensurate
transition and still remaining gapped at the transition point.

In other words ferromagnetic legs favour the magnon attraction and in
this case an incommensurate phase is exotic. However this phase is favorable for
antiferromagnetic legs with repulsion of magnons.

This general argumentation may be confirmed by exact results obtained
for two exactly solvable models. The first one was studied in
\cite{5}. The corresponding Hamiltonian is the following:
\begin{eqnarray}
{\cal H}&=&\sum_nJ_{\bot}{\bf S}_{1,n}{\bf S}_{2,n}+J_{\|}({\bf
S}_{1,n}{\bf S}_{1,n+1}+{\bf S}_{2,n}{\bf
S}_{2,n+1}\nonumber\\
&+&4({\bf S}_{1,n}{\bf S}_{1,n+1})({\bf S}_{2,n}{\bf
S}_{2,n+1}))-g\mu_Bh({\bf S}_{1,n}+{\bf S}_{2,n}),
\end{eqnarray}
where ${\bf S}_{i,n}$ ($i=1,2;\,n=-\infty...\infty$) are
spin-$\frac{1}{2}$ operators associated with cites of the ladder.

As it was shown in \cite{5} the corresponding values of critical
and saturation magnetic fields are
\begin{equation}
g\mu_Bh_c=J_{\bot}-4J_{\|},\qquad g\mu_Bh_s=J_{\bot}+4J_{\|},
\end{equation}
It follows from (5) that $J_{\|}<0$ entails $h_s<h_c$, so the
incommensurate phase does not appear.

Another solvable model of spin-ladder with cyclic (ring) exchange
and exact singlet-rung ground state is considered in the next
section.

\section{Magnetic behavior of a spin-ladder with exact singlet-rung ground state}
The model of spin-ladder with cyclic exchange and exact
singlet-rung ground state in magnetic field $h$ was first
suggested in \cite{11} and studied in more detail in \cite{12}.
The corresponding Hamiltonian $\cal H$ has the following form:
\begin{equation}
{\cal H}=\sum_{n=-\infty}^{\infty}H_{n,n+1},
\end{equation}
where
$H_{n,n+1}=H^{stand}_{n,n+1}+H^{frust}_{n,n+1}+H^{cyc}_{n,n+1}+H^{Zeeman}_{n,n+1}+J_{norm}$,
and
\begin{eqnarray}
H^{stand}_{n,n+1}&=&\frac{J_{\bot}}{2}({\bf S}_{1,n}\cdot{\bf
S}_{2,n}+{\bf S}_{1,n+1}\cdot{\bf S}_{2,n+1})+ J_{\|}({\bf
S}_{1,n}\cdot{\bf S}_{1,n+1}+{\bf S}_{2,n}\cdot{\bf
S}_{2,n+1}),\nonumber\\
H^{frust}_{n,n+1}&=&J_{frust}({\bf S}_{1,n}\cdot{\bf
S}_{2,n+1}+{\bf S}_{2,n}\cdot{\bf
S}_{1,n+1}),\nonumber\\
H^{cyc}_{n,n+1}&=&J_c(({\bf S}_{1,n}\cdot{\bf S}_{1,n+1})({\bf
S}_{2,n}\cdot{\bf S}_{2,n+1})+({\bf S}_{1,n}\cdot{\bf
S}_{2,n})({\bf S}_{1,n+1}\cdot{\bf S}_{2,n+1})\nonumber\\
&-&({\bf S}_{1,n}\cdot{\bf S}_{2,n+1})({\bf S}_{2,n}\cdot{\bf
S}_{1,n+1})),\nonumber\\
H^{Zeeman}_{n,n+1}&=&-g\mu_B({\bf S}^3_{1,n}+{\bf S}^3_{2,n}).
\end{eqnarray}
The constant term $J_{norm}$ is added only for normalization to
zero the ground state energy at $h=0$.

As it was shown in \cite{11} the following restriction:
\begin{equation}
J_{frust}=J_{\|}-\frac{1}{2}J_c,
\end{equation}
guarantees that the Hamiltonian (6),(7) commutes with the
triplon-number operator ${\cal Q}=\frac{1}{2}\sum_{n}({\bf
S}_{1,n}+{\bf S}_{1,n+1})^2$. In this case the Hilbert space
separates on sectors corresponding to different eigenvalues of
${\cal Q}=0,1,2,...$. The sector ${\cal Q}=0$ is generated by the
single vector (1). The additional restrictions:
\begin{equation}
J_{norm}=\frac{3}{4}J_{\bot}-\frac{9}{16}J_c,\quad
J_{\bot}>2J_{\|},\quad J_{\bot}>\frac{5}{2}J_c,\quad
J_{\bot}+J_{||}>\frac{3}{4}J_c,
\end{equation}
guarantee that the state (1) is the zero-energy ground state of
the Hamiltonian (6),(7) separated by a finite gap from the other
states.

The following formulas corresponding to zero magnetic field,
\begin{eqnarray}
H_{n,n+1}|0\rangle_n|1\rangle_{n+1}^j&=&
(\frac{1}{2}J_{\bot}-\frac{3}{4}J_c)|0\rangle_n|1\rangle_{n+1}^j+\frac{J_c}{2}|1\rangle_n^j|0\rangle_{n+1},\nonumber\\
H_{n,n+1}|1\rangle_n^j|0\rangle_{n+1}&=&
(\frac{1}{2}J_{\bot}-\frac{3}{4}J_c)|1\rangle_n^j|0\rangle_{n+1}+\frac{J_c}{2}|0\rangle_n|1\rangle_{n+1}^j,\\
H_{n,n+1}|1\rangle_n^1|1\rangle_{n+1}^1&=&(J_{\bot}+J_{\|}-\frac{3}{4}J_c)|1\rangle_n^1|1\rangle_{n+1}^1,\\
H_{n,n+1}|1\rangle_n^0|1\rangle_{n+1}^1&=&(J_{\bot}-\frac{J_c}{2})|1\rangle_n^0|1\rangle_{n+1}^1+
(J_{\|}-\frac{J_c}{4})|1\rangle_n^1|1\rangle_{n+1}^0,\nonumber\\
H_{n,n+1}|1\rangle_n^1|1\rangle_{n+1}^0&=&(J_{\bot}-\frac{J_c}{2})|1\rangle_n^1|1\rangle_{n+1}^0+
(J_{\|}-\frac{J_c}{4})|1\rangle_n^0|1\rangle_{n+1}^1,
\end{eqnarray}
are useful for calculation of magnon dispersions.

If we suppose that the system does not pass through the
incommensurate phase then the energy gap between the states (1)
and (3) will be closed at the saturation field $h_s$. Then as it
follows from (11)
\begin{equation}
g\mu_Bh_s=J_{\bot}+J_{\|}-\frac{3}{4}J_c.
\end{equation}

The simplest one-magnon excitation of the ground state (1) has the following "triplon" form:
$\\|k\rangle_{trip}^j=\sum_n
e^{ikn}...|0\rangle_{n-1}|1\rangle_n^j|0\rangle_{n+1}...\,$. It
has the following dispersion
$E_{triplon}(k)=J_{\bot}-3/2J_c+J_c\cos k$ \cite{11}. Positivity
of the one-triplon gap at $h=h_s$ entails the following condition:
\begin{equation}
E^{trip}_{gap}(h_s)=-(J_{\|}+\frac{3}{4}J_c+|J_c|)>0.
\end{equation}
From (14) follows that $J_{\|}<0$ irrespective to $J_c$.

Excitations near the state $|vac\rangle_s$ are also gapped at
$h=h_s$. We shall calculate only the gaps corresponding to $\Delta
S^z=-1$ sectors with minimal increase of magnetic energy.

There are two types of these states. The first one "singlon"
$\\|k\rangle_{sing}=\sum_n{\rm
e}^{ikn}...|1\rangle^1_{n-1}|0\rangle_n|1\rangle^1_{n+1}...$,
originates from "annihilations" of rung-triplets into the
corresponding rung-singlets. Standard calculation based on
(10),(11) gives the following dispersion law for these states:
$E^{singl}(k)=-J_{\|}-3/4J_c+J_c\cos k$. The corresponding energy
gap coincides with (14).

Excitation of the second type "ferromagnon":
$|k\rangle_{fmagn}=\sum_n{\rm
e}^{ikn}...|1\rangle^1_{n-1}|1\rangle_n^0|1\rangle^1_{n+1}...$,
originates from rotations of single rung-triplets. Its dispersion
$E^{fmagn}(k)=J_{\bot}-J_{\|}-J_c/4+2(J_{\|}-J_c/4)\cos k$ may be
easily calculated from (11),(12) and corresponds to the following
gap:
\begin{equation}
E^{fmagn}_{gap}(h_s)=J_{\bot}-J_{\|}-J_c/4-2|J_{\|}-J_c/4|.
\end{equation}

It may be easily proved using (15) and (9) that
$E^{sing}_{gap}<E^{fmagn}_{gap}$. This fact has a clear
interpretation. In both the cases an increase of magnetic energy
is equal. However singlons have an additional energy decrease
originated from the triplet-singlet annihilation. On the other
hand ferromagnons have an additional energy increase caused by the
destruction of the ferromagnetic order. By this reason the energy
gap corresponding to ferromagnons must be bigger than the one
corresponding to singlons.

\section{Magnetic behavior at various temperatures}

It was argued that for spin-ladders with ferromagnetic legs an increase of external magnetic field
inspires direct first order transition between two gapped phases: non-magnetic singlet-rung and
ferromagnetic fully polarized.
This transition occurs without passing throw a gapless incommensurate phase. In other words the transition interval
between $h_c$ and $h_s$ turns to zero for these materials. Here we discuss some
characteristic features of the corresponding experimental behavior.

Of course the step-like jump
of magnetization will be effectively noticeable only for $T<E_{gap}(h_s)$.
For $T>E_{gap}(h_s)$ contributions from excited states will smooth the magnetic curve getting it
similar to the one corresponding to passing through an incommensurate phase. For these temperatures it will be
difficult to distinguish a spin ladder with low $E_{gap}(h_s)$ from the one with low $h_s-h_c$.

Since the effective difference between magnetic curves for spin-ladders with weak ferromagnetic and
antiferromagnetic legs disappears for $T>E_{gap}(h_s)$ it is nesessary to use some other
experimental approaches to study the nature of these compounds.
Probably a good criterion may be obtained from an independent
measuring of the $E_{gap}$ (from magnetic susceptibility, neutron or Raman scattering data e.t.c.).
If at low temperatures the magnetic curve reduces to the step-like form as well as it was previously exactly
determined that $E_{gap}>h_c$ then we have a right to content that a nonmagnetic gapped phase
turns directly into a gapped ferromagnetic. We suggest that this fact also indicates a ferromagnetism along legs.

It was reported in \cite{9} that the spin-ladder compound
$(5{\rm IAP})_2{\rm CuBr}_4\cdot2{\rm H}_2{\rm O}$ has a weak antiferromagnetic interaction
along legs ($J_{||}/J_{\bot}=0.077$). Its magnetic behavior is very similar to the one discussed
above. Really the magnetization curve at $T=0.4$ K
has a pronounced step-like form and almost vertical slope compared
to the $T=4.35$ K case. This kind of behavior drastically
differs from the one corresponding to another spin-ladder compound
${\rm MgV}_2{\rm O}_5$ \cite{8} where the slope of the magnetic
curve does not change significantly under a decrease of
temperature.

Analysis of the temperature dependence of magnetic susceptibility for $(5{\rm IAP})_2{\rm CuBr}_4\cdot2{\rm H}_2{\rm O}$
reported in \cite{9} gives $E_{gap}=h_c=12.23$ K, however an analysis of the
magnetization curve gives $h_c=11.90$ K. If we suppose that theses results indicate a weak ferromagnetism along legs
then we have to conclude that at the transition point the system has an extremely little gap
$E_{gap}(h_s)=0.33$ K. Of course this conjecture may be confirmed only in a more precise experiment at $T<0.33 K$.

\section{Conclusions}

In this paper we have suggested that spin-ladders with
ferromagnetic legs have different magnetic behavior than the ones
with antiferromagnetic legs. We have confirmed our qualitative
arguments by considerations of two exactly solvable models.
This effect will be really noticeable only at low temperatures ($T\ll E_{gap}(h_s)$).
For rather big temperatures the real difference of magnetization curves for spin-ladders with
weak ferromagnetic or antiferromagnetic legs must be negligible.
We also have noticed that the spin-ladder compound $(5{\rm
IAP})_2{\rm CuBr}_4\cdot2{\rm H}_2{\rm O}$ may be discussed in
this context.

Spin-ladders with ferromagnetic legs have a special theoretical
interest (see \cite{6} and references therein). However the
characteristic type of their magnetic behavior suggested in the
present paper was not previously discussed.

Field-induced first order phase transitions in spin-ladders were also studied in \cite{7}.
However the corresponding phase diagram was different from the one suggested in
the present paper. In \cite{7} non-magnetic and fully polarized phases are
separated by the phase with half polarization.

The author is very grateful to Xi-Wen Guan for the interesting
discussion.

\end{document}